\documentclass[preprint,aps]{revtex4}

\newcommand {\bea}{\begin{eqnarray}}
\newcommand {\eea}{\end{eqnarray}}
\newcommand {\be}{\begin{equation}}
\newcommand {\ee}{\end{equation}}

\newcommand {\Dslash}{D\!\!\!/}

\newcommand {\fq}{(qq)(\bar{q}\bar{q})}
\newcommand {\tqq}{(q\bar{q})^2}
\newcommand {\tq}{(q\bar{q})}
\newcommand {\gf}{\gamma_5}

\usepackage{graphicx}

\begin{document}


\title{Instantons and Scalar Multiquark States: From Small
to Large $N_c$}

\author{T.~Sch\"afer$^{1,2}$}

\affiliation{
$^1$Department of Physics, North Carolina State University,
Raleigh, NC 17695\\
$^2$Riken-BNL Research Center, Brookhaven National 
Laboratory, Upton, NY 11973}

\begin{abstract}
  We study scalar $(\bar{q}q)$ and $(\bar{q}q)^2$ correlation
functions in the instanton liquid model. We show that the 
instanton liquid supports a light scalar-isoscalar (sigma) meson, 
and that this state is strongly coupled to both  $(\bar{q}q)$ and 
$(\bar{q}q)^2$. The scalar-isovector $a_0$ meson, on the other 
hand, is heavy. We also show that these properties are specific
to QCD with three colors. In the large $N_c$ limit the 
scalar-isoscalar meson is not light, and it is mainly 
coupled to $(\bar{q}q)$.

\end{abstract}
\maketitle

\section{Introduction}
\label{sec_intro}

 There is a longstanding  controversy concerning the structure of
scalar-isoscalar mesons in QCD \cite{Spanier:du,Jaffe:1976ig,Weinstein:gc,Lohse:ew,Tornqvist:1995ay,Black:1998wt,Alford:2000mm,Isgur:2000ts,Schafer:2000hn}.
One view is that scalar mesons are $p$-wave $q\bar{q}$ bound states 
with masses in the (1.0-1.5) GeV range. According to this view the large 
effects seen in the $I=0$ $\pi\pi$ and $I=1/2$ $\pi K$ channel are due 
to $t$-channel meson exchanges and other coupled channel effects. The 
competing view is that there are genuine scalar resonances below 1 GeV,
and that the structure of these states is quite different from ordinary 
quark model $q\bar{q}$ states. 

 Jaffe suggested that the unusual properties of the light scalar
mesons could be explained by assuming a large $(qq)(\bar{q}\bar{q})$
admixture \cite{Jaffe:1976ig,Alford:2000mm}. He observed that the
spectrum of the flavor nonet obtained by coupling two anti-triplet 
scalar diquarks is inverted as compared to a standard $q\bar{q}$ nonet, 
and contains a light isospin singlet, a strange doublet, and a heavy 
triplet plus singlet with hidden strangeness. This compares very favorably
to the observed light sigma, the strange kappa, and the heavier
$a_0(980)$ and $f_0(980)$. It also explains why the $a_0$ and $f_0$ 
are strongly coupled to $K\bar{K}$ and $\pi\eta$. Strong correlations 
in the anti-triplet scalar diquark channel are quite natural in QCD 
and have also been invoked in order to understand the structure of 
baryons \cite{Anselmino:1992vg,Schafer:1993ra} and dense baryonic 
matter \cite{Rapp:1997zu,Alford:1997zt}.

 In this work we study the nature of scalar-isoscalar mesons
in the instanton model \cite{Diakonov:1995ea,Schafer:1996wv}.
There are several reasons why instantons effects are relevant.
Scalar-isoscalar mesons have vacuum quantum numbers and are 
directly related to fluctuations in the chiral condensate. There 
is an impressive amount of evidence from lattice calculations 
that instantons are responsible for chiral symmetry breaking 
in QCD \cite{DeGrand:2000gq,Faccioli:2003qz,Horvath:2001ir}.
These calculations also suggest that the approximate zero
modes responsible for chiral symmetry breaking are small 
in size. This implies that the chiral condensate is very 
inhomogeneous and that fluctuations in the scalar channel 
are large. In addition to that, if there is a large $(qq)
(\bar{q}\bar{q})$ admixture in the wave function of scalar 
mesons, both flavor mixing and off-diagonal $(q\bar{q})-
(q\bar{q})^2$ mixing have to be large. The instanton 
induced $(q\bar{q})^{N_f}$ interaction provides a natural
mechanism for these effects. 

 In order to study the nature of scalar-isoscalar mesons
we compute both $(q\bar{q})$ and $(qq)(\bar{q}\bar{q})$ 
correlation functions. We show that there is a large 
scalar resonance signal in both the diagonal $\tq$ and 
$\fq$ correlators as well as the off-diagonal $\tq$-$\fq$ 
correlator. This implies that there is indeed a large
four quark admixture in the scalar-isoscalar meson. 
We have to note, however, that in a relativistic
quantum field theory this statement is somewhat 
hard to quantify. We can compute the coupling of 
the $\tq$ and $\fq$ operators to the scalar resonance, 
but these coupling constant have different dimensions
and it is not clear how to compare them. 

 We address this problem by studying scalar correlators
for different numbers of colors. We expect that in the
large $N_c$ limit the scalar meson becomes a ``normal''
quark model state which is dominated by the $\tq$ 
component. This implies that there is no scalar resonance
signal in the four quark channel, and that the 
scalar-isoscalar and scalar-isovector correlators 
become indistinguishable. We show that this tendency 
is clearly seen in ratios of correlation functions 
computed in QCD with different numbers of colors. 

 This paper is organized as follows. In Sect.~\ref{sec_cor} 
we define $\tq$ and $\fq$ scalar correlation functions. In 
Sect.~\ref{sec_inst} we discuss the short distance behavior 
of these correlators and investigate the importance of 
unphysical contributions in the quenched approximation. In
Sect.~\ref{sec_num} we present numerical results from quenched 
and unquenched simulations of the instanton liquid model. 

\section{Scalar correlators}
\label{sec_cor}

\begin{figure}
\leavevmode
\begin{center}
\includegraphics[width=15cm,angle=0]{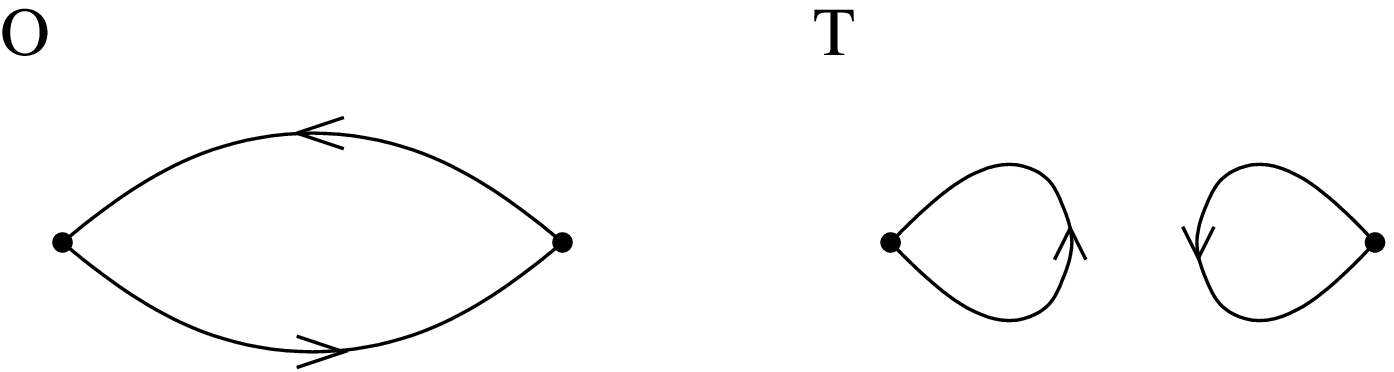}
\end{center}
\caption{\label{fig_sigcor}
Quark line diagrams contributing to the scalar $\tq$
 correlation function, ($O$) ``one-loop'', and  
($T$) ``two-loop''. The lines are quark propagators 
in a gluonic background field.}
\end{figure}

 The lowest dimension operator with the quantum numbers
of the sigma meson is the scalar $\tq$ operator
\be
\label{j_sig}
j_{\tq} = \frac{1}{\sqrt{2}} (\bar{u}u+\bar{d}d) .
\ee
The corresponding correlation function is (see 
Fig.~\ref{fig_sigcor})
\be
\label{pi_sig}
\Pi_{\tq}(x) = O(x)-2T(x)
 =\langle {\rm Tr}\left[S(0,x)S(x,0)\right]\rangle 
 -  2\langle {\rm Tr}\left[S(0,0)\right]
   {\rm Tr}\left[S(x,x)\right]\rangle ,
\ee
where $\langle.\rangle$ is an average over all gluonic field
configurations, $S(0,x)$ is the full quark propagator in a 
given background gluonic field, and the trace is taken over
Dirac and color indices. $O(x)$ and $T(x)$ denote the one-loop 
(connected) and two-loop (disconnected) terms in the correlation 
function. Note that the quark line connected term is equal to 
the correlator of the scalar-isovector current, $\Pi_{a_0}(x)
=O(x)$. Lattice calculations of the correlation function 
equ.~(\ref{pi_sig}) have been reported in 
\cite{Michael:1999rs,Muroya:2001yp,Prelovsek:2002qs,Kunihiro}.

\begin{figure}
\leavevmode
\begin{center}
\includegraphics[width=15cm,angle=0]{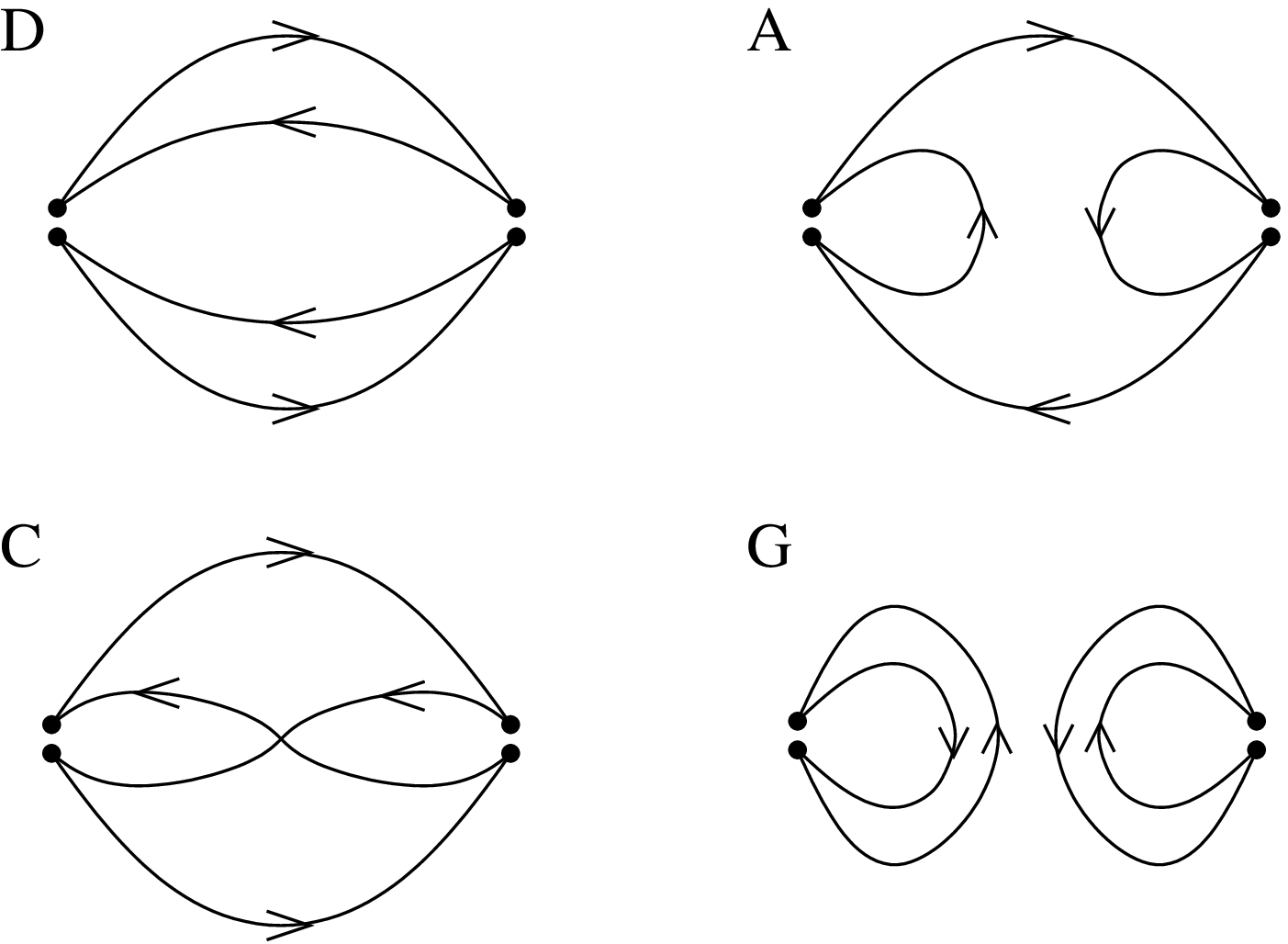}
\end{center}
\caption{\label{fig_cor}
Quark line diagrams contributing to the two-pion correlation 
function, ($D$) ``direct'', ($C$) ``crossed'', ($A$) ``single 
annihilation'', and  ($G$) ``double annihilation'' or ``glue''. }
\end{figure}

\begin{figure}
\leavevmode
\begin{center}
\includegraphics[width=15cm,angle=0]{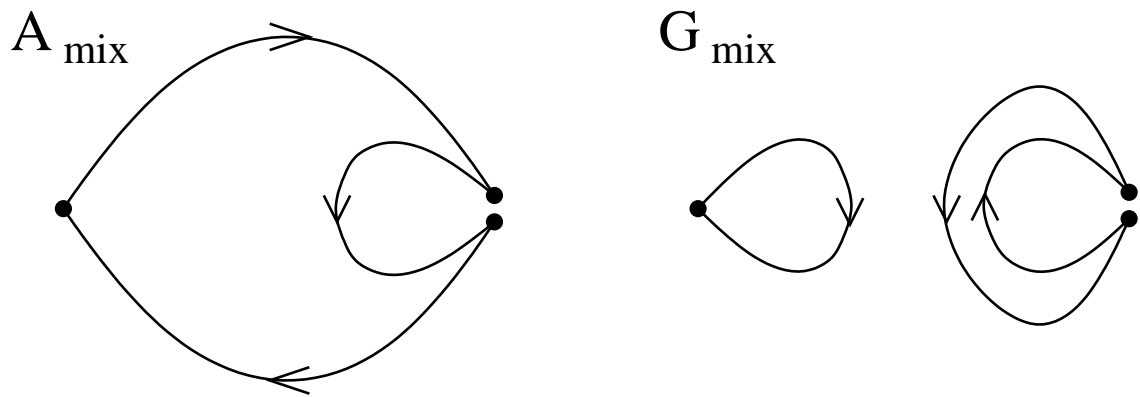}
\end{center}
\caption{\label{fig_mix}
Quark line diagrams contributing to the off-diagonal scalar 
two-pion correlation function, ($A_{mix}$) ``single annihilation'', 
and ($G_{mix}$) ``double annihilation'' or ``glue''. }
\end{figure}

 There are several ways to construct scalar-isoscalar 
four quark operators. One possibility is to couple two
quark-anti-quark operators. We expect the most attractive 
channel to arise from the coupling of two color-singlet,
flavor-triplet pseudoscalar operators,
\be 
\label{j_i0}
j_{\tqq}= (\bar{q}\gf\tau^a q) (\bar{q}\gf\tau^a q).
\ee
The corresponding correlation function is equal to 
the $I=0$ two-pion correlator \cite{Gupta:1993rn}
\be 
\label{pi_i0}
\Pi_{I0}(x) =
\frac{1}{4!} \langle j_{\tqq}(0)j_{\tqq}(x)\rangle 
 =  D(x)+\frac{1}{2}C(x) - 3A(x) +\frac{3}{2} G(x),
\ee
where
\bea
\label{pid}
 D(x) &=& \langle \left({\rm Tr}\left[S(0,x)\gf S(x,0)\gf
  \right]\right)^2\rangle  \\
 C(x) &=& \langle  {\rm Tr}\left(\left[S(0,x)\gf S(x,0) \gf
  \right]^2\right)\rangle  \\
 A(x) &=& \langle {\rm Tr}\left[S(0,x)\gf S(x,x)\gf S(x,0) \gf
 S(0,0)\gf  \right]\rangle  \\
 G(x) &=& \langle {\rm Tr}\left[\left(S(0,0)\gf \right)^2
  \right] {\rm Tr}\left[\left(S(x,x)\gf \right)^2
  \right]\rangle
\eea
are the four quark correlators shown in Fig.~\ref{fig_cor}.
The $I=2$ two-pion correlator is given by 
\be
\label{pi_i2}
\Pi_{I2}(x)= D(x)-C(x)
\ee 
and does not involve any disconnected diagrams. The
mixed $\tq$-$\tqq$ correlation function is given by (see
Fig.~\ref{fig_mix})
\be 
\label{pi_mix}
\Pi_{\tq-I0}(x) = 
\frac{1}{3!} \langle j_{\tq}(0)j_{\tqq}(x)\rangle 
 = \sqrt{2}\left( A_{mix}(x)-G_{mix}(x)\right),
\ee
where 
\bea
 A_{mix}(x) &=& \langle  {\rm Tr}\left[S(0,x)\gf S(x,x)
   \gf S(x,0) \right] \rangle  \\
 G_{mix}(x) &=& \langle {\rm Tr}\left[ S(0,0) \right] 
  {\rm Tr}\left[\left(S(x,x)\gf \right)^2
  \right]\rangle .
\eea
Instead of coupling two quark-anti-quark operators we
can construct scalar-isoscalar four quark operators by 
coupling a diquark and an anti-diquark operator. The 
most attractive channel is expected to arise from two
color anti-triplet and isospin singlet scalar diquarks
\be
\label{dq2}
 j_{(dq)^2} = \epsilon^{abc}\epsilon^{ade}
  (q^bC\gf\tau_2 q^c)(\bar{q}^dC\gf\tau_2\bar{q}^e).
\ee
This operator can be Fierz-rearranged into a $\tqq$
operator. The result is
\bea
\label{dqf}
 j_{(dq)^2} &=& \frac{1}{4}\Big\{
 (\bar{q}\gf\bar{\tau}^a q)(\bar{q}\gf\bar{\tau}^a q)
 +(\bar{q}\bar{\tau}^a q)(\bar{q}\bar{\tau}^a q) 
 +(\bar{q}\gamma_\mu\bar{\tau}^a q)
  (\bar{q}\gamma_\mu\bar{\tau}^a q)
 \nonumber \\
 & & \hspace{1.6cm}\mbox{}
 +(\bar{q}\gf\gamma_\mu\bar{\tau}^a q)
  (\bar{q}\gf\gamma_\mu\bar{\tau}^a q)
 -\frac{1}{2}(\bar{q}\sigma_{\mu\nu}\bar{\tau}^a q)
   (\bar{q}\sigma_{\mu\nu}\bar{\tau}^a q)
  \Big\}
\eea
where $\bar{\tau}^a=(\vec{\tau},i)$. Note that the flavor 
structure of equ.~(\ref{dqf}) is that of a determinant, 
$(\bar{q}\bar{\tau}^a q)(\bar{q}\bar{\tau}^a q)=(\bar{u}u)
(\bar{d}d)-(\bar{u}d)(\bar{d}u)$. Equation (\ref{dqf}) contains 
both attractive channels, such as $\pi\pi$ and $f_0f_0$, as
well as repulsive channels, such as $\eta'\eta'$ and $a_0a_0$. 
In the following we will concentrate on the $\pi\pi$ channel 
equ.~(\ref{j_i0}).

\section{Scalar correlators in the instanton model}
\label{sec_inst}

 In this section we wish to study instanton contributions to the scalar 
correlation functions equ.~(\ref{pi_sig},\ref{pi_i0},\ref{pi_i2})
and equ.~(\ref{pi_mix}). At short distances only the effect of 
the closest instanton has to be taken into account. The main
instanton contribution is related to the zero mode term
in the fermion propagator
\be
\label{S_SIA}
S(x,y) \simeq \frac{\psi_{0}(x-z)\psi^\dagger_{0}(y-z)}{m^*},
\ee
where $\psi_0(x)$ is the zero mode wave function and $z$ is
the location of the instanton. For an isolated instanton 
$m^*=m_q$ where $m_q$ is the current quark mass. In an 
ensemble of instantons $m^*$ is an effective mass $m^*\simeq 
\pi\rho(2/N_c)^{1/2} (N/V)^{1/2}$. Here, $\rho$  is the average 
instanton size and $N/V$ is the average instanton density. 
Instanton zero modes make equal contributions to the one and 
two-loop terms in the scalar correlation function equ.~(\ref{pi_sig}), 
$O_{inst}=T_{inst}<0$. As a consequence the one-instanton 
contribution is attractive in the $\sigma$ channel and repulsive 
in the $a_0$ channel. Averaging over the position of the instanton
we get
\be
\label{ps_SIA}
\Pi^{SIA}_{\sigma,a_0}(x) = \pm\int d\rho\, n(\rho)
 \frac{6\rho^4}{\pi^2}\frac{1}{(m^*)^2}
 \frac{\partial^2}{\partial (x^2)^2}
 \left\{ \frac{4\xi^2}{x^4} \left(
 \frac{\xi^2}{1-\xi^2} +\frac{\xi}{2}
 \log\frac{1+\xi}{1-\xi}\right)\right\},
\ee
where $n(\rho)$ is the instanton size distribution and
$\xi^2=x^2/(x^2+4\rho^2)$. The one-instanton contribution can 
be resummed using the random phase approximation (RPA) 
\cite{Diakonov:1985eg,Hutter:1994av,Kacir:1996qn}
\be
\label{pi_rpa}
\Pi^{RPA}_{\sigma,a_0}(x) = N_c \left(\frac{N_cV}{N}\right)
  \int d^4q\, e^{iq\cdot x}\;\;
           \Gamma_s(q)\frac{\pm 1}{1\pm C_s(q)}\Gamma_s(q) .
\ee
The loop and vertex functions $C_s$ and $\Gamma_s$ are given by
\bea
\label{s_bubble_mfa}
 C_s (q) &=& 4N_c\left(\frac{V}{N}\right)
  \int \frac{d^4p}{(2\pi)^4}
  \frac{M_1M_2(M_1M_2-p_1\cdot p_2)}
       {(M_1^2+p_1^2)(M_2^2+p_2^2)} ,\\
\label{s_vertex_mfa}
\Gamma_s (q) &=& \;\; 4  \int \frac{d^4p}{(2\pi)^4}
  \frac{(M_1M_2)^{1/2}(M_1M_2-p_1\cdot p_2)}
       {(M_1^2+p_1^2)(M_2^2+p_2^2)}  ,
\eea
where $p_1=p+q/2$, $p_2=p-q/2$ and $M_{1,2}=M(p_{1,2})$
are momentum dependent effective quark masses. In the 
instanton liquid model the momentum dependence of the 
effective quark mass is governed by the Fourier transform 
of the instanton zero mode profile. The effective quark 
mass at zero virtuality, $M(0)$, is determined by a 
Dyson-Schwinger equation. For typical values of the 
parameters $\rho\simeq 0.3$ fm and $(N/V)=1\,{\rm fm}^{-4}$ 
we find $M(0)\simeq 350$ MeV. 

\begin{figure}
\leavevmode
\begin{center}
\includegraphics[width=15cm,angle=0]{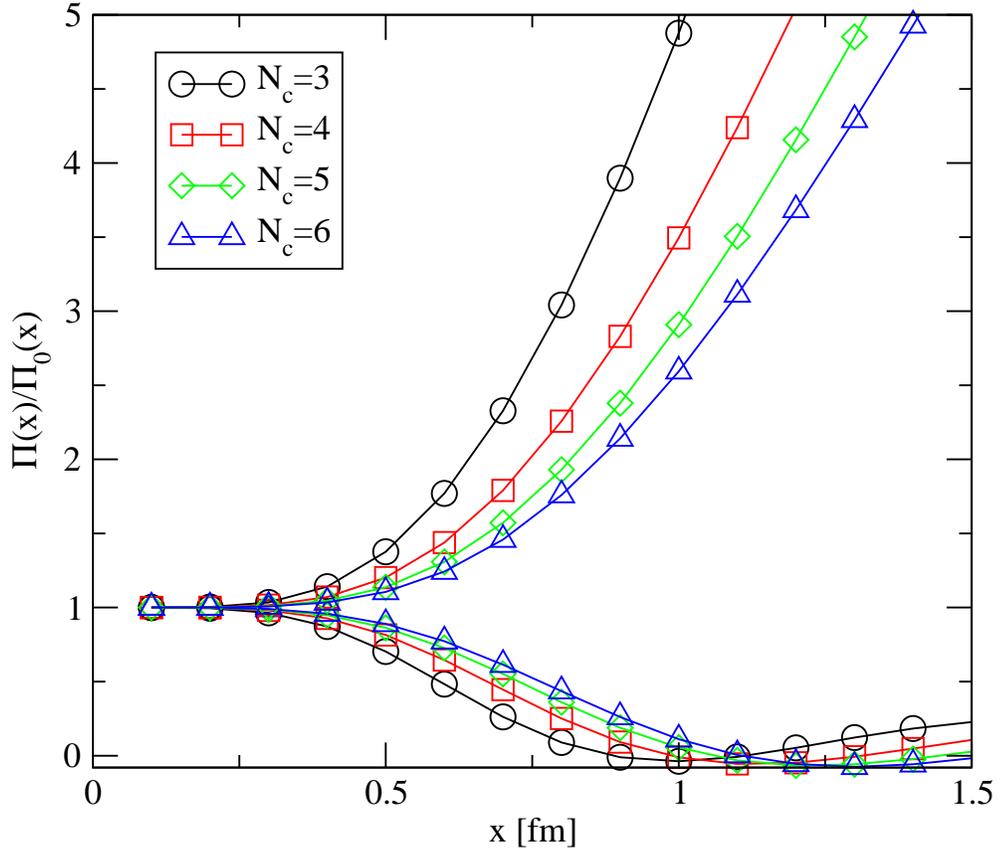}
\end{center}
\caption{\label{fig_mfa}
Scalar-isoscalar ($\sigma$) and scalar-isovector ($a_0$)
correlation functions for different numbers of colors
$N_c$ calculated in the RPA approximation to the 
instanton liquid model. The correlation functions 
are normalized to free field behavior $\Pi_0(x)\sim
1/x^6$. The upper/lower curves correspond to the 
$\sigma/a_0$, respectively.}
\end{figure}

 Scalar correlation functions in the RPA approximation
are shown in Fig.~(\ref{fig_mfa}). We observe that the 
isoscalar and isovector correlation functions are indeed 
very different. We also note that the isovector correlator
is unphysical for $x>0.7$ fm. This is an artifact of 
the quenched approximation and can be understood in 
terms of quenched $\pi\eta'$ intermediate states. The 
quenched $\sigma$ correlation function receives unphysical 
$\eta'\eta'$ contributions, but these terms are not included 
in the random phase approximation. The scalar-isoscalar 
($\sigma$) correlator shows a clear resonance signal. A 
simple pole fit gives $m_\sigma=540$ MeV. We observe that 
the $\sigma$ and $a_0$ correlation functions become more 
similar as the number of colors increases. However, in 
the RPA approximation the difference remains finite and 
large in the limit $N_c\to\infty$. As explained in 
\cite{Schafer:2002af} this is related to a failure of 
the  RPA approximation when applied to the large $N_c$ 
instanton liquid. 

 Let us now study instanton contributions to the two-pion 
correlation functions equ.~(\ref{pi_i0}) and (\ref{pi_i2}). 
The first possibility is that all four quarks propagate
in zero mode states. In QCD with $N_f=2,3$ flavors 
this contribution violates the Pauli principle and has 
to vanish. This is reflected by the fact that in both 
the $I=0$ and $I=2$ correlation functions the sum of the 
coefficients of the four different contractions, $D,C,A$ 
and $G$, vanishes. We note that for light quark masses 
it is dangerous to drop the disconnected contributions to 
the $I=0$ correlator, an approximation sometimes referred 
to as the quenched valence approximation. In particular, 
there is a large positive zero mode contribution to the 
connected $I=0$ correlation function which is not present 
in the full $I=0$ correlator.

\begin{figure}
\leavevmode
\begin{center}
\includegraphics[width=15cm,angle=0]{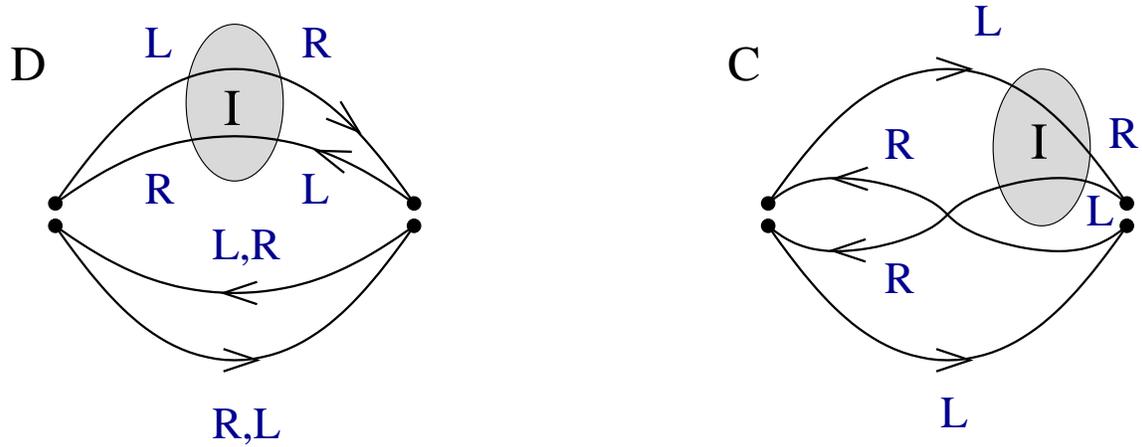}
\end{center}
\caption{\label{fig_pipi_sia}
Instantons contributions to the $\pi\pi$ correlation functions. 
At the one-instanton level, there are no contributions to the 
single and double annihilation diagrams.}
\end{figure}

 Contributions with an odd number of fermion zero modes violate
chirality. This leaves terms with two zero mode propagators,
see Fig.~\ref{fig_pipi_sia}. These terms only appear in 
the direct and crossed diagrams. The one-instanton 
contribution to the direct diagram is $D^{SIA}(x)=
2\Pi^0_\pi(x)\Pi^{SIA}_\pi(x)$, where $\Pi^0_\pi(x)=
3/(\pi^4x^6)$ is the free quark contribution and 
$\Pi^{SIA}_\pi(x)=\Pi^{SIA}_\sigma(x)$ is the single
instanton contribution to the pion correlation function. 
It is clear that $D^{SIA}(x)$ simply corresponds to two 
non-interacting pions. Interactions arise from the crossed 
term. We find
\bea
\Pi^{SIA}_{I0}(x)&=&2\Pi^0_\pi(x)\Pi^{SIA}_\pi(x)
  + \frac{1}{6}\Pi^0_\pi(x)\Pi^{SIA}_\pi(x) \\
\Pi^{SIA}_{I2}(x)&=&2\Pi^0_\pi(x)\Pi^{SIA}_\pi(x)
  -\frac{1}{3}\Pi^0_\pi(x)\Pi^{SIA}_\pi(x).
\eea
This implies that the one-instanton term is attractive
in the $I=0$ channel and repulsive in the $I=2$ channel. 
One can try to resum this interaction in order to study 
the $\pi\pi$ interaction at intermediate and long distances. 
For the one-gluon exchange interaction, this problem was 
recently studied in \cite{Cotanch:2002vj}. At very low 
momenta, the $\pi\pi$ interaction is constrained by 
chiral symmetry. Weinberg predicted the $s$-wave 
scattering lengths in the $I=0$ and $I=2$ channels 
\cite{Weinberg:1966kf}
\be
a_0^0=\frac{7m_\pi}{32\pi f_\pi^2},\hspace{1cm}
a_0^2=-\frac{m_\pi}{16\pi f_\pi^2}.
\ee
We emphasize, however, that our objective in this 
work is not a determination of the scattering lengths, 
but a study of scalar resonances. In Sect.~\ref{sec_num} 
we will describe how to sum the instanton induced interaction 
to all orders by performing numerical simulations of 
the $I=0,2$ $\pi\pi$ correlation functions in the 
instanton model. 

At the one-instanton level there is no contribution 
to the the mixed $\tq$-$\tqq$ correlation 
function. This is a consequence of the fact that 
in $N_f=2$ QCD not all quarks can propagate in zero
mode states. In QCD with three flavors there is a
flavor mixing $(\bar{u}u)(\bar{d}d)(\bar{s}s)$ 
contribution.

\begin{figure}
\leavevmode
\begin{center}
\includegraphics[width=15cm,angle=0]{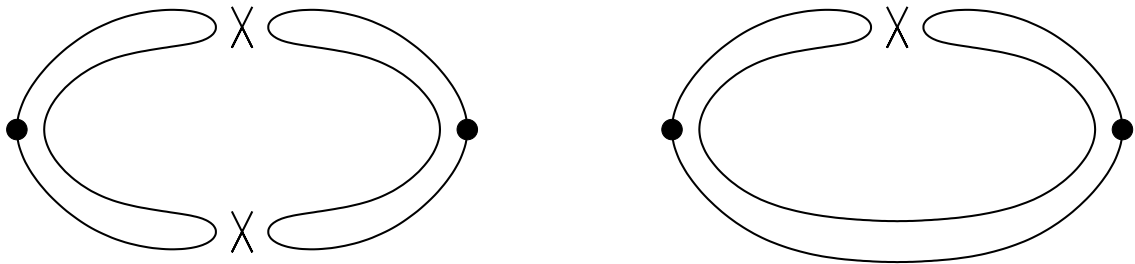}
\end{center}
\caption{\label{fig_hair}
Diagrams that lead to quenching artifacts in the two-pion
correlation function. The diagram on the left and right
show the two and one double-hairpin terms.}
\end{figure}

\begin{figure}
\leavevmode
\begin{center}
\includegraphics[width=15cm,angle=0]{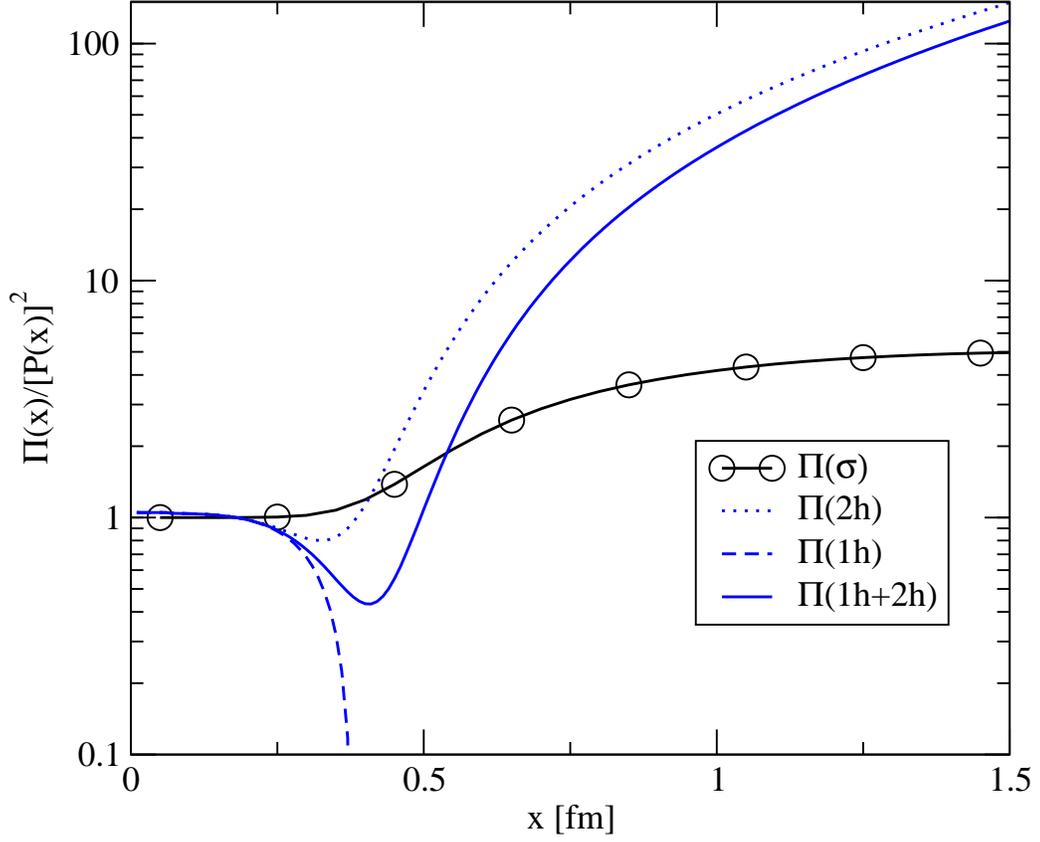}
\end{center}
\caption{\label{fig_etaeta}
Quenched $\eta'\eta'$ contribution to the $I=0$ 
$\pi\pi$ correlation function normalized to the 
square of the pion correlation function. The 
curves labeled (1h) and (2h) show the one and 
two double-hairpin diagram, see Fig.~\ref{fig_hair}.
For comparison, we also show the contribution
of a scalar resonance.}
\end{figure}

 Finally, we have to understand the effect of the quenched
approximation on the isoscalar two-pion correlation function
\cite{Bernard:zp,Bernard:1995ez}, see Fig.~\ref{fig_hair}.
The main artifact in the quenched approximation arises
from unphysical $\eta'\eta'$ intermediate states. The 
long distance, low momentum part of the quenched $\eta'$
propagator is of the form
\be
\label{eta_qu}
\Pi_{\eta'}(q) = \frac{\lambda_\pi^2}{q^2+m_\pi^2}
   -\frac{\lambda_\pi^2}{q^2+m_\pi^2}m_0^2
    \frac{1}{q^2+m_\pi^2},
\ee
where the second term is the so-called double-hairpin 
contribution, $m_0^2\simeq m_{\eta'}^2$ is related
to the $\eta'$ mass and $\lambda_\pi$ is the coupling
of the pion to the pseudoscalar current. Note that the 
double hairpin term is negative in euclidean space. 
Using equ.~(\ref{eta_qu}) we see that the $\eta'\eta'$
contribution to the $I=0$ $\pi\pi$ correlation function
contains a one double-hairpin term which is negative 
and a two double-hairpin term which is positive in 
euclidean space. Numerical results for the coordinate space 
correlation function are shown in Fig.~\ref{fig_etaeta}.
The correlation function is normalized to the square 
of the pion correlator. For comparison we also show 
the contribution of a scalar $\sigma$ resonance with 
mass $m_\sigma=500$ MeV and coupling constant $\lambda_\sigma
=(350\,{\rm MeV})^5$. We observe that the quenched 
approximation leads to unphysical contributions that 
are repulsive at intermediate distances $x\sim 0.5$
fm and large and attractive at large distance. We also
observe that these effects will mask the presence of 
a sigma resonance, unless the sigma coupling is very 
large.

\section{Numerical Results}
\label{sec_num}

 In this section we present numerical results obtained in
quenched and unquenched simulations of the instanton 
liquid model \cite{Schafer:1995pz,Shuryak:1992ke}. The main 
assumption of the instanton liquid model is that the QCD 
partition function
\be 
 Z= \int DA_\mu \,\exp(-S)\prod_f^{N_f} 
  \det(i\Dslash+im_f)
\ee
is dominated by classical gauge configurations called 
instantons. Instantons in QCD with $N_c$ colors are 
characterized by $4N_c$ collective coordinates, position
(4), size (1), and color orientation ($4N_c-5$). An 
ensemble of instantons is governed by the partition
function
\be 
 Z= \sum_{N_I,N_A}\frac{1}{N_I\!N_A\!}
\int \prod_i^{N_I+N_A}\left[d\Omega_id(\rho_i)\right]\,
  \exp(-S)\prod_f^{N_f} 
  \det(i\Dslash+im_f),
\ee
where $\Omega=(z_i,\rho_i,U_i)$ labels the collective 
coordinates and $d(\rho)$ is the instanton measure.
Correlation functions are computed from the quark 
propagator in a given instanton configuration. The 
propagator is determined by numerically inverting the 
Dirac operator in the space of approximate zero modes.
The short distance part is added perturbatively. 

 Most of the results presented below were obtained 
from simulations at a quark mass $m_q=20$ MeV in a 
euclidean box with dimensions $(2.83\,{\rm fm})^3\times 
5.66\,{\rm fm}$. We have not systematically studied 
finite volume corrections. Finite volume effects 
on the two-pion correlator can be used in order to
determine pion scattering lengths \cite{Luscher:1985dn}.
Our aim in this work is not a determination of the 
$\pi\pi$ scattering length, but a study of $\pi\pi$
resonances.

\begin{figure}
\leavevmode
\begin{center}
\includegraphics[width=15cm,angle=0]{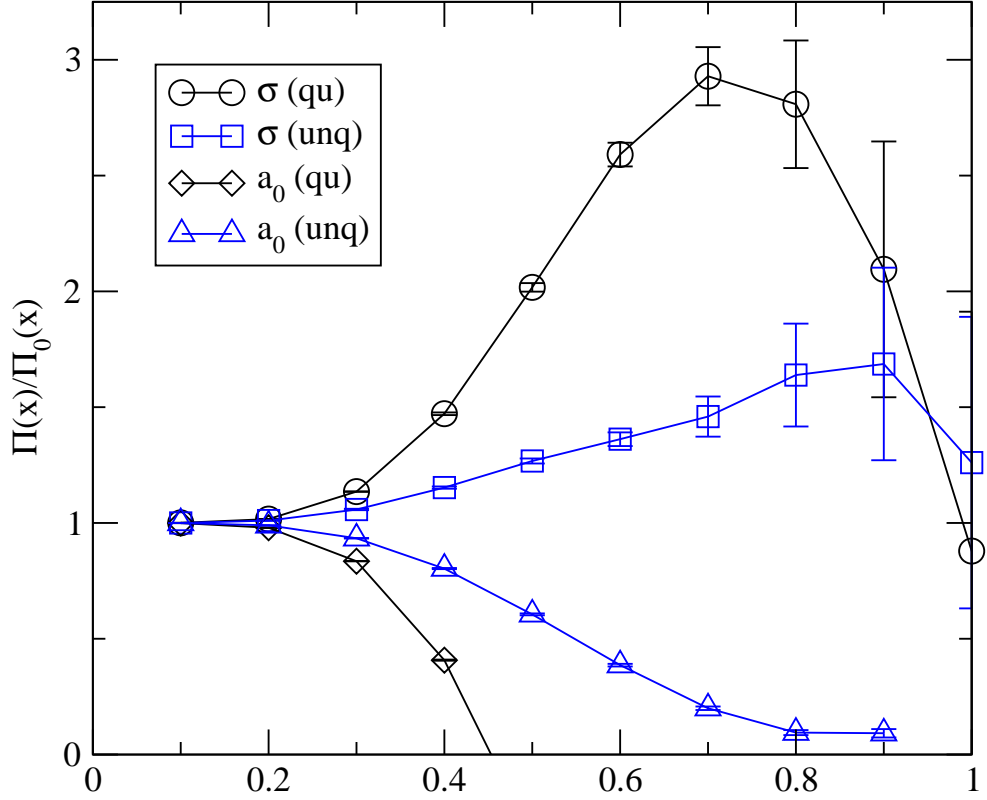}
\end{center}
\caption{\label{fig_sig_del}
Scalar-soacalar $\sigma$ and scalar-isovector $a_0$
correlation functions in the instanton model. The 
correlators are normalized to free field behavior 
$\Pi_0(x)\sim 1/x^6$. The labels (qu) and (unq) 
refer to the quenched and unquenched correlation
functions.}
\end{figure}

 Scalar-isoscalar and scalar-isovector correlation
functions are shown in Fig.~\ref{fig_sig_del}. We
note that the scalar-isoscalar $\sigma$ and 
scalar-isovector $a_0$ correlators behave very
differently. The $a_0$ correlator is very repulsive.
In the quenched approximation the correlation 
function even becomes negative. This behavior 
can be understood in terms of quenched $\eta'\pi$ 
intermediate states. In full QCD the correlator is
still quite repulsive. We extract a resonance mass
$m_{a_0}\sim 1$ GeV. The sigma correlator, on the
other hand, is very attractive, both in quenched 
and full QCD. Because of the large vacuum contribution
$\langle(\bar{q}q)^2\rangle$ it is hard to extract
the mass. Our results are consistent with a light 
scalar $m_\sigma=(0.55-0.70)$ GeV.

\begin{figure}
\leavevmode
\begin{center}
\includegraphics[width=15cm,angle=0]{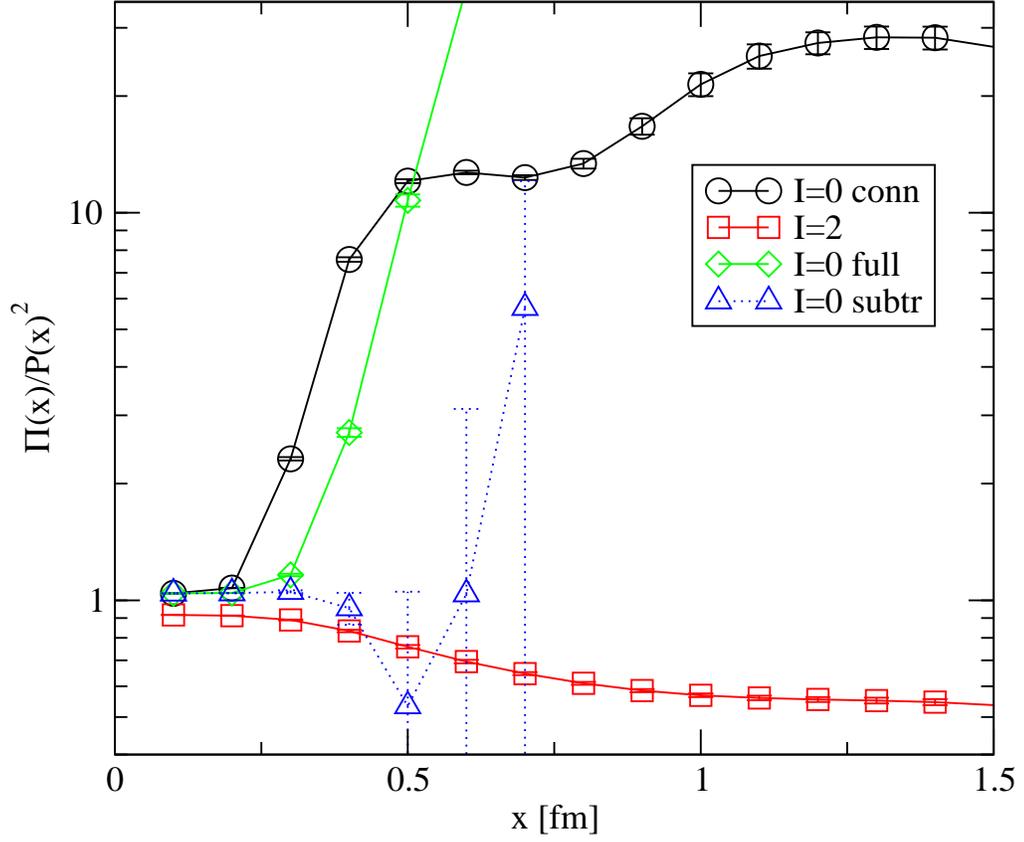}
\end{center}
\caption{\label{fig_i0i2_qu}
$I=0$ and $I=2$ two-pion correlation functions measured
in quenched simulations of the instanton liquid model.
The correlators are normalized to the square of the 
pion correlation function. The curve labeled (conn) 
shows the connected part of the $I=0$ correlator, 
(full) is the complete $I=0$ correlator, and (subtr) 
is the complete correlator after the vacuum 
contribution is subtracted.}
\end{figure}

 In Fig.~\ref{fig_i0i2_qu} we show $I=0,2$ $\pi\pi$ 
correlation functions in the quenched approximation. 
The results are normalized to the square of the pion
correlation function. We observe that the $I=2$ $\pi\pi$
correlator is indeed repulsive, the correlation function
is suppressed as compared to the square of the pion
correlator. The connected part of the $I=0$ correlator 
is very strongly enhanced and has an unusual shape. This 
is due to the effect discussed in Sect.~\ref{sec_inst}.
The connected $I=0$ correlator receives a large 
Pauli principle violating zero mode contribution. 
The full $I=0$ correlator has a vacuum contribution 
$\langle (\bar{q}\tau^a q)^2\rangle^2$ that needs
to be subtracted. The subtraction involves large 
cancellations and the result has large statistical 
uncertainties for $x>0.5$ fm. We note that the 
subtracted correlation function is repulsive at
intermediate distances $x\sim 0.5$ fm. As explained
in Sect.~\ref{sec_inst} this is an artifact of the 
quenched approximation. 

\begin{figure}
\leavevmode
\begin{center}
\includegraphics[width=15cm,angle=0]{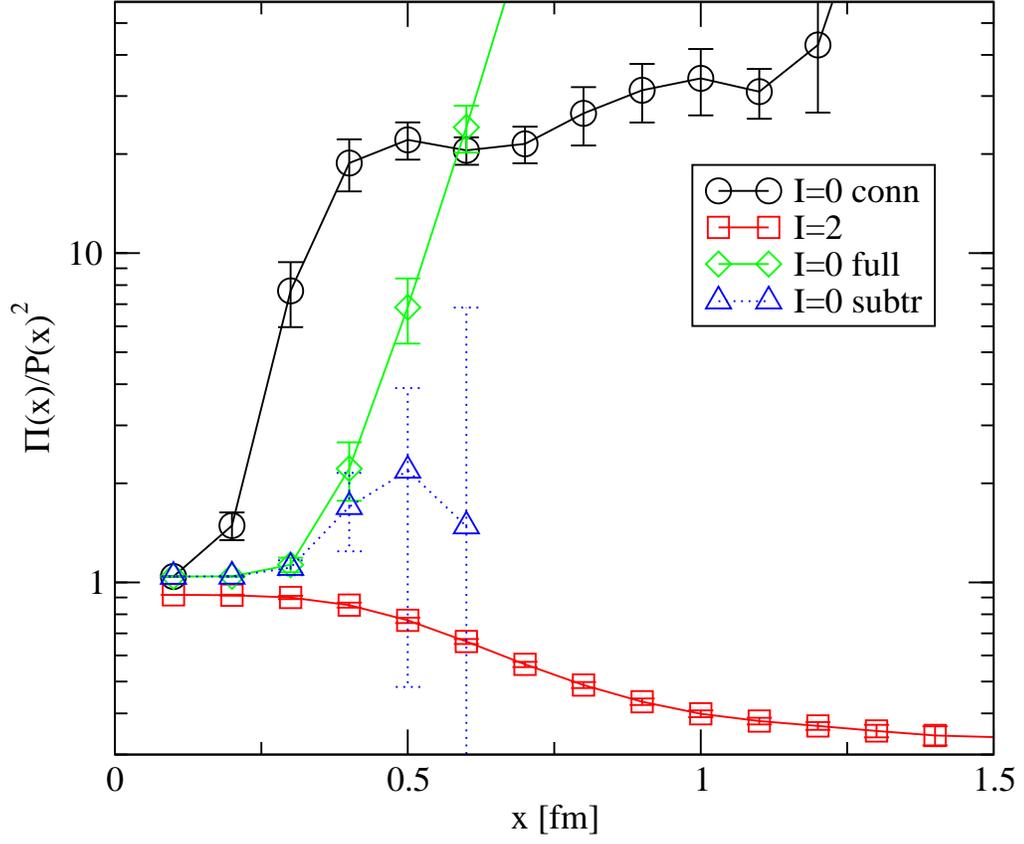}
\end{center}
\caption{\label{fig_i0i2_unq}
$I=0$ and $I=2$ two-pion correlation functions  measured
in unquenched simulations of the instanton liquid model.
Curves labeled as in Fig.~\ref{fig_i0i2_qu}.}
\end{figure}

 Unquenched $\pi\pi$ correlation functions are shown
in Fig.~\ref{fig_i0i2_unq}. We observe that the 
$I=2$ correlation function is not strongly modified
as compared to the unquenched calculation. The same
is true for the connected $I=0$ correlator. However, 
the full (subtracted) $I=0$ correlator is strongly 
modified. The correlation function is now attractive 
at intermediate distances $x\sim 0.5$ fm. The data 
are not sufficiently accurate to determine the mass
of a resonance, but the resonance parameters can 
be estimated if the data are combined with the diagonal
$\tq$ and off-diagonal $\tq-\tqq$ correlators. We 
find $m_\sigma=600$ MeV and $\lambda_\sigma=(330\, {\rm
MeV})^5$.

\begin{figure}
\leavevmode
\begin{center}
\includegraphics[width=15cm,angle=0]{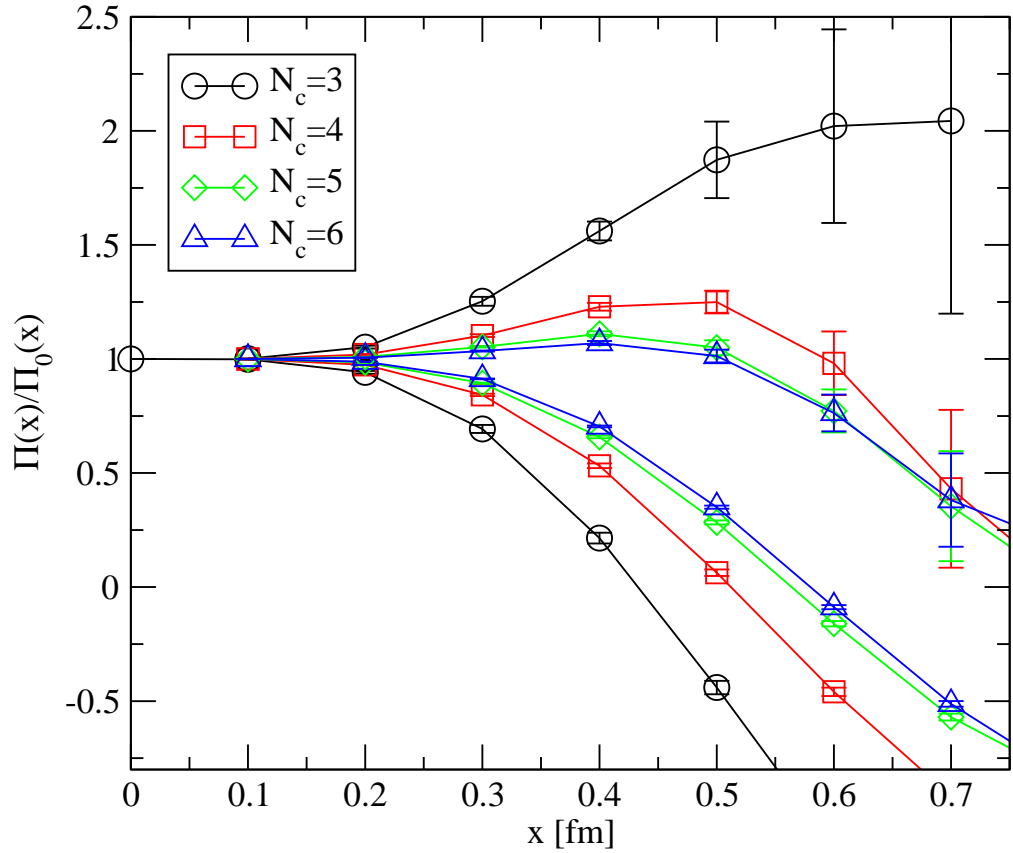}
\end{center}
\caption{\label{fig_scal_nc}
Scalar-isoscalar ($\sigma$) and scalar-isovector ($a_0$)
correlation functions for different numbers of colors
$N_c$. The correlators are normalized to free field 
behavior, $\Pi^0(x)\sim N_c/x^6$. The upper/lower curves 
correspond to the $\sigma/a_0$, respectively. The data 
shown in this figure were obtained from quenched simulations 
of the instanton liquid model.}
\end{figure}

\begin{figure}
\leavevmode
\begin{center}
\includegraphics[width=15cm,angle=0]{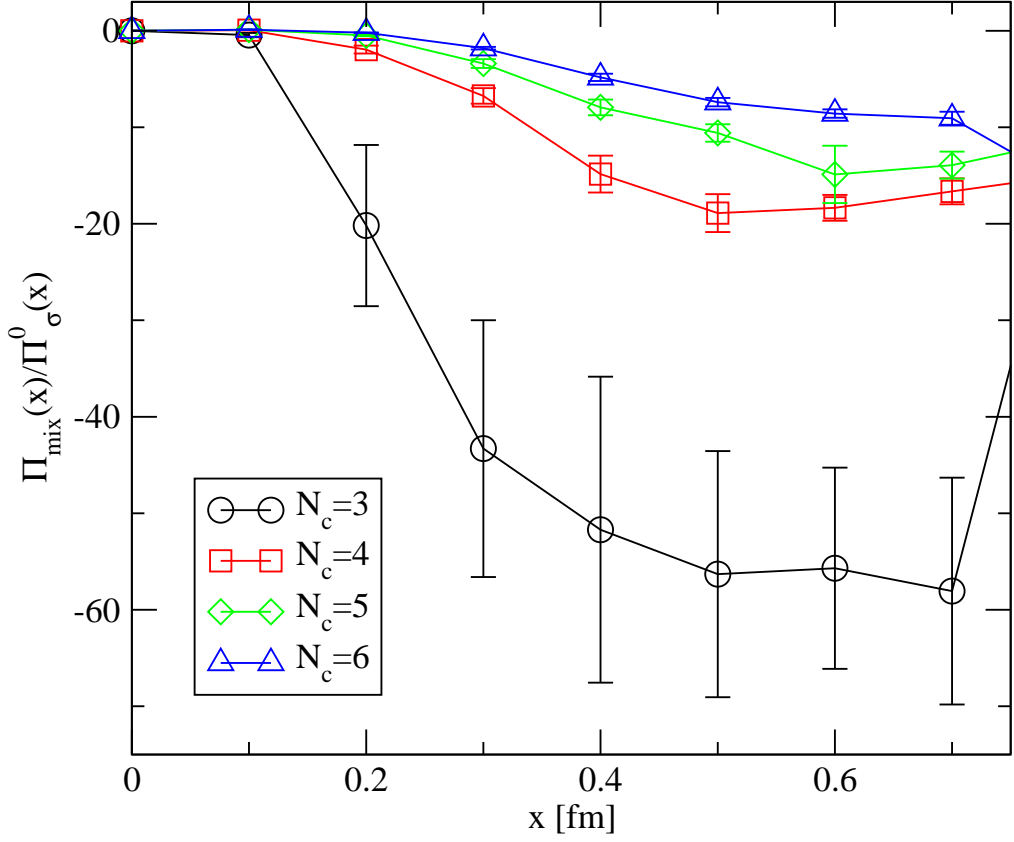}
\end{center}
\caption{\label{fig_mix_nc}
Off-diagonal $\tq$-$\tqq$ correlation function 
for different values of the numbers of colors $N_c$.
The correlation functions are normalized to the 
free $\tq$ correlator. The data shown in 
this figure were obtained from quenched simulations 
of the instanton liquid model.}
\end{figure}

\begin{figure}
\leavevmode
\begin{center}
\includegraphics[width=15cm,angle=0]{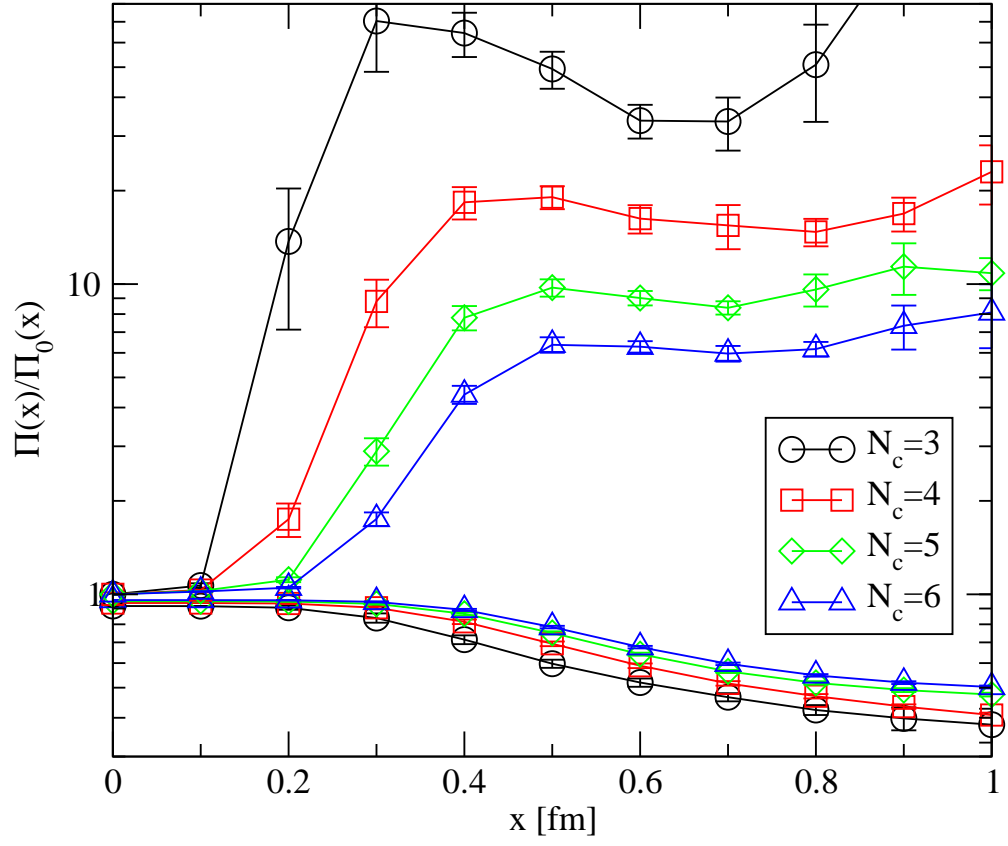}
\end{center}
\caption{\label{fig_i0i2_nc}
Connected $I=0$ and $I=2$ two-pion correlation functions
for different values of the numbers of colors $N_c$.}
\end{figure}

 In Figs.~\ref{fig_scal_nc}-\ref{fig_i0i2_nc} we
show the behavior of the correlation functions for
different numbers of colors $N_c=3,\ldots,6$ 
\cite{Schafer:2002af}. We find that the difference
between the scalar-isoscalar and scalar-isovector 
correlation functions disappears as the number 
of colors increases. For $N_c=6$ the mass of the 
sigma is $m_\sigma>1 $ GeV. The physical reason
for the change in the sigma meson correlation function
is the fact that the chiral condensate becomes more 
homogeneous as the number of colors increases. The 
main change in the scalar-isovector correlation function 
is the disappearance of quenched $\eta'\pi$ intermediate
states. We expect that the unquenched $a_0$ mass is
fairly independent of the number of colors. 

  Fig.~\ref{fig_mix_nc} shows that the off-diagonal 
$\tq-\tqq$ correlation function decreases dramatically 
as the number of colors increases. The results are 
consistent with the idea that the off-diagonal 
correlation function vanishes in the large $N_c$ 
limit. The $I=2$ $\pi\pi$ correlator is only weakly 
affected by the number of colors. We observe, however,
that the $I=2$ correlation function tends towards the 
non-interacting $\pi\pi$ correlator. The connected 
$I=0$ correlator, on the other hand, shows a very 
dramatic decrease towards the non-interaction two 
pion correlation function as the number of colors
increases.

\section{Summary}
\label{sec_sum}

 In this paper we studied scalar $\tq$ and $\tqq$ 
correlation functions in the instanton model. We showed 
that the structure of the scalar-isoscalar $\sigma$ and 
scalar-isovector $a_0$ mesons is very different. The 
correlation function in the $a_0$ channel is very repulsive 
while the $\sigma$ channel is strongly attractive. The 
scalar $\tq$ correlator is consistent with a light sigma 
resonance with a mass $m_\sigma=(550-700)$ MeV 
\cite{Shuryak:1992ke,Schafer:1995pz,Hutter:1994av,Schafer:2000hn}. 
A light sigma resonance also appears in Nambu-Jona-Lasinio 
models \cite{Hatsuda:1994pi}. However, the large OZI violation 
in the scalar channel is characteristic of the instanton induced 
interaction.

 We also showed that the sigma couples strongly to 
the $I=0$ $\tqq$ current. Observing this effect 
not only requires a calculation of both the connected 
and disconnected diagrams, but also the inclusion
of the fermion determinant. If the calculation is 
restricted to connected diagrams, or performed in 
the quenched approximation, the scalar resonance
is masked by large unphysical contributions.

 We showed that the special properties of the sigma 
meson, its small mass and large coupling to $\tqq$ 
states, are specific to QCD with three colors. If the 
number of colors is increased the sigma becomes an 
``ordinary'' meson with a mass in the 1 GeV range and 
small coupling to $\tqq$ states. The physical reason is 
that the chiral condensate is very inhomogeneous
in $N_c=3$ QCD, but becomes more homogeneous as the
number of colors increases. The large inhomogeneity 
of the condensate in $N_c=3$ QCD reflects the small 
size of the instanton. The chiral condensate becomes 
more homogeneous as $N_c$ grows because the instanton 
liquid is more dense. As emphasized in \cite{Schafer:2002af}
this does not necessarily imply that instantons 
overlap strongly, since the instanton density grows 
like the volume of the color group.

 There are many interesting problems that remain to be 
studied. In this paper we concentrated on scalar mesons 
in $N_f=2$ flavor QCD. One of the remarkable aspects of 
$\fq$ states in $N_f=3$ QCD is their flavor structure. 
In particular, one would like to verify that there is 
hidden strangeness in the heavy scalars. In addition to 
the meson-meson or diquark-anti-diquark interaction we 
would also like to study diquark-diquark correlations. 
This problem is relevant to the structure of dense baryonic 
matter \cite{Rapp:1999qa}, the spectrum of $(qq)^2\bar{q}$ 
pentaquark states \cite{Jaffe:2003sg}, and the possible 
existence of the H-dibaryon \cite{Jaffe:1976yi}.
 
Acknowledgments: We would like to thank E.~Shuryak for useful 
discussions. This work was supported in part by US DOE grant 
DE-FG02-03ER41260.


\newpage 

\begin{figure}
\leavevmode
\begin{center}
\includegraphics[width=15cm,angle=0]{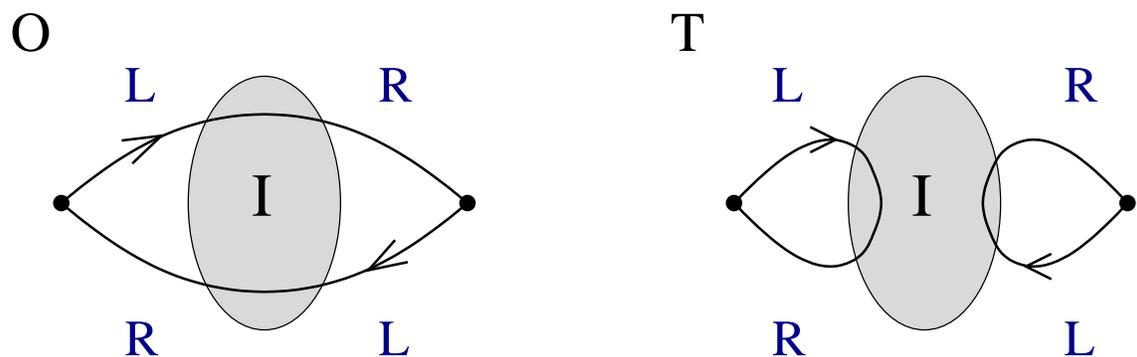}
\end{center}
\caption{\label{fig_sig_sia}
Single instanton contributions to the scalar correlation function.
The lines crossing the shaded area correspond to zero mode 
propagators, and $L,R$ label the chirality of the quark. The
anti-instanton contribution corresponds to $L\leftrightarrow R$. }
\end{figure}

\end{document}